\documentclass[%
 twocolumn,
 amsmath,amssymb,
 aps,
 floatfix,
]{revtex4-2}

\usepackage{graphicx}
\usepackage{dcolumn}
\usepackage{bm}
\usepackage{subfigure}
\usepackage{color}
\usepackage{ulem}
\newcommand{\red}[1]{{\color{black} #1}}

\begin{document}


\title{Random sequential adsorption of aligned regular polygons and rounded squares: Transition in the kinetics of packing growth}
\author{Micha\l{} Cie\'sla}%
 \email{michal.ciesla@uj.edu.pl}
\author{Piotr Kubala}%
 \email{piotr.kubala@doctoral.uj.edu.pl}
\affiliation{Institute of Theoretical Physics, Jagiellonian University, 30-348 Krak\'ow, \L{}ojasiewicza 11, Poland}%
\author{Aref Abbasi Moud}
\email{aabbasim@ucalgary.ca}
\affiliation{Department of Chemical and Biological Engineering, The University of British Columbia, Vancouver, British Columbia V6T 1Z3, Canada}

\date{\today}

\begin{abstract}
We study \red{two-dimensional} random sequential adsorption \red{(RSA)} of \red{flat} polygons and rounded squares \red{aligned in parallel} to find a transition in the asymptotic behavior of the kinetics of packing growth. Differences in the kinetics for RSA of disks and parallel squares were confirmed in previous analytical and numerical reports. Here, by analyzing the two classes of shapes in question we \red{can} precisely control the shape of packed figures and thus localize the transition. Additionally, we study how the asymptotic properties of the kinetics depend on the packing size. We also provide accurate estimations of saturated packing fractions. The microstructural properties of generated packings are analyzed in terms of the density autocorrelation function. 
\end{abstract}

\maketitle

%
%
%
\section{Introduction}
Random sequential adsorption (RSA) is a numerical protocol used for generating random packings \cite{Evans1993, Talbot2000, Kubala2022}. According to it, the shapes are placed randomly one after another, however, the placing occurs only if the next shape does not overlap any of the previously added shapes. After placing, the position and orientation of each figure \red{remain} unchanged. The procedure continues until the packing is saturated -- there is no place for any other shape. In contrast to the so-called random close packings (RCP) where the neighboring particles typically are in contact \cite{Torquato2010}, here the packing is rather loose and the mean packing fraction is significantly smaller. 

Although the history of RSA begins in 1939 when Flory used \red{the random process described above} to study the structure of a linear polymer to which some groups of molecules can be attached at random places \cite{Flory1939}, the real interest in RSA began in 1980 when Feder noticed that the structure of such two-dimensional random packings resembles monolayers produced in irreversible adsorption experiments \cite{Feder1980}. The similarities were so \red{substantial} that saturated packing fractions of disks on a flat surface were determined using adsorption experiments \cite{Onoda1986}. On the other hand, the numerical generation of large, strictly saturated packings was ineffective because when the packing is almost saturated, the probability that a randomly placed and oriented object will not intersect with any previously added one is tiny. Thus, the number of such attempts has to be very large to place the next figure. Although for some specific shapes, there exist methods overcoming this problem e.g. \cite{Wang1994, Haiduk2018, Ciesla2018}, the properties of saturated state are \red{still} often estimated using the kinetics of packing growth \red{computed for} almost saturated packings. For a majority of shapes, the asymptotic kinetics is given by Feder's law
\begin{equation}
\label{eq:fl}
    \theta - \theta(t) \sim t^{-\frac{1}{d}},
\end{equation}
when the packing is close to a saturated state. Here, $\theta$ is the saturated packing fraction, and $\theta(t)$ is the packing fraction after $t$ tries of adding a shape to the packing. Parameter $d$ depends on shape and packing \red{dimensionality}. For example, for \red{$k$-dimensional (hyper)spheres packed in the (hyper)space of the same dimensionality, $d = k$}, while for anisotropic, randomly oriented \red{two-dimensional} shapes placed on the \red{two-dimensional} flat surface $d=3$ \cite{Vigil1989, Viot1992, Haiduk2018}.
On the other hand, for parallel squares or rectangles 
\begin{equation}
\label{eq:log}
    \theta - \theta(t) \sim \frac{\log t}{t}.
\end{equation}
Both these relations were confirmed analytically \cite{Pomeau1980, Swendsen1981} and numerically e.g.~\cite{Vigil1989, Brosilow1991, Viot1992, Shelke2007, Haiduk2018}

Here, we want to study the transition between these two regimes. We tried to achieve this in two ways. The first one is to generate \red{two-dimensional} random packings composed of \red{flat} regular polygons aligned in parallel. For \red{the} RSA of squares, as noted above, the kinetics of packing growth is governed by (\ref{eq:log}). When the number of regular polygon sides grows, its shape approaches the disk for which kinetics is given by (\ref{eq:fl}) with parameter $d=2$. A similar study was recently presented in \cite{Moud2022examination}, but the author focused on saturated packing fractions while the presented results on the RSA kinetics of \red{squares} did not agree with the analytical law (\ref{eq:log}). The second way is to generate packings built of aligned squares with rounded corners. By increasing the radius of this rounding the shape approaches the disk, thus the transition in the packing's growth kinetics should be visible. The second method seems to be superior to the first one because the radius can be changed continuously, while the number of regular polygon's sides is a discrete value and the disk is approached only in the limit of \red{an} infinite number of sides.
\section{Numerical details}
RSA protocol consists of iterations of the following steps:
\begin{itemize}
    \item select the position of a virtual polygon randomly with the probability uniformly distributed over the packing;
    \item check if the virtual particle does not overlap with any polygon inside the packing;
    \item if it does not, add it to the packing, otherwise, remove and abandon it.
\end{itemize}
To generate strictly saturated \red{packings} according to RSA protocol we traced the regions where subsequent particles can be added. This idea was used for the first time by Akeda et al. in the case of packings built of parallel squares \cite{Akeda1975} and by Wang for disks \cite{Wang1994}. The method is based on the division of the packing into small regions called voxels, and each voxel is tested if there is \red{a} possibility to place there \red{the} center of the next shape without overlapping existing polygons. If not, such a voxel is removed \red{from the list of existing voxels}. Thus, the random sampling of the position of the virtual shape is limited only to \red{the voxels that are on the list}, which speeds up the packing generation. The voxels can be divided into smaller ones to better estimate the region when placing is possible. The simulation ends where there are no voxels left, thus, the packing is saturated. A variant of this method for polygons was invented by Zhang \cite{Zhang2018} and improved further in \cite{Ciesla2019b} and the details about the voxel removal criterion can be found there. Although in its original version, this method was designed for the generation of saturated packings built of arbitrarily oriented polygons, its restriction to a single orientation is straightforward.

It should be mentioned that when the sampling of the virtual shape position covers only existing voxels occupying a fraction of the whole packing surface area $S$, one iteration corresponds to $S/S_v$ iterations in the original RSA protocol, where $S_v$ is the total surface area of these voxels. Additionally, to compare the results obtained for different sizes of packings, the number of iterations is expressed in the so-called dimensionless time units where one unit contains $S/S_p$ iterations. Here, $S_p$ is the surface area of a single polygon. Throughout this study, the number of iterations shall be expressed in these units and denoted as $t$.

We studied saturated RSA packings built of regular polygons of the number of sides ranging from $3$ to $1000$. To estimate the kinetics and other properties, we generated $100$ independent random packings for each type of polygon. The figures were placed on the square of the surface area $S=10^6$, while the surface area of a single polygon was normalized to $S_p=1$. To minimize finite-size effects, periodic boundary conditions were used \cite{Ciesla2018}. The number of iterations needed to form saturated packing differs significantly between independent packings, as \red{it} in general \red{is} distributed according to a heavy tail probability distribution function \cite{Ciesla2017}. Therefore, to estimate the asymptotic value of the parameter $d$ in the power law (\ref{eq:fl}) we restricted to the data from the range $[t_\text{min}/100, t_\text{min}]$, where $t_\text{min}$ is the dimensionless time when the first packing becomes saturated. Such an approach guarantees sufficient statistics and also analyzes kinetics close to saturation. \red{It is worth noting that $\log [\theta(2t) - \theta(t)]$ exhibits the same asymptotic scaling as (\ref{eq:fl})  when plotted against $\log(t)$ \cite{Baule2017,Pooria2019}, which gives another way to determine the exponent $d$.}

\begin{figure}
    \centering
    \includegraphics[width=0.7\columnwidth]{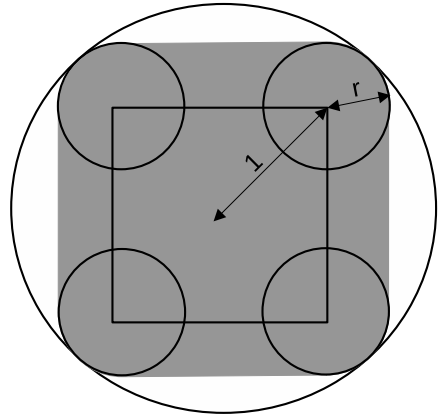}
    \caption{Parametrization of a rounded square. The circumscribed circle of the square has a unit radius and circumscribed circle of the rounded square has a radius \red{of} $1 + r$.}
    \label{fig:roundedSqure}
\end{figure}

\red{Similarly we studied RSA packings of rounded squares.} The shape is parameterized by one additional parameter $r$ which corresponds to the circle radius at each corner of the square  -- see Fig.~\ref{fig:roundedSqure}.
\red{Parameter $r$ can vary from $0$ (square) to infinity (disk), but here it was restricted to $r \in [0,1]$. The surface area of the shape in Fig.~\ref{fig:roundedSqure} is $S_{r} = 2 + 4 r\sqrt{2} + \pi r^2$, and the linear size of the rounded square was always rescaled to obtain $S_{r} = 1$.} This \red {parameterization} was used in Ref.~\cite{Ciesla2021}, where one can also find a detailed description of the method for generating saturated RSA packing built of rounded polygons.
\section{Results}
\begin{figure}[ht]
    \centering
    \includegraphics[width=0.4\columnwidth]{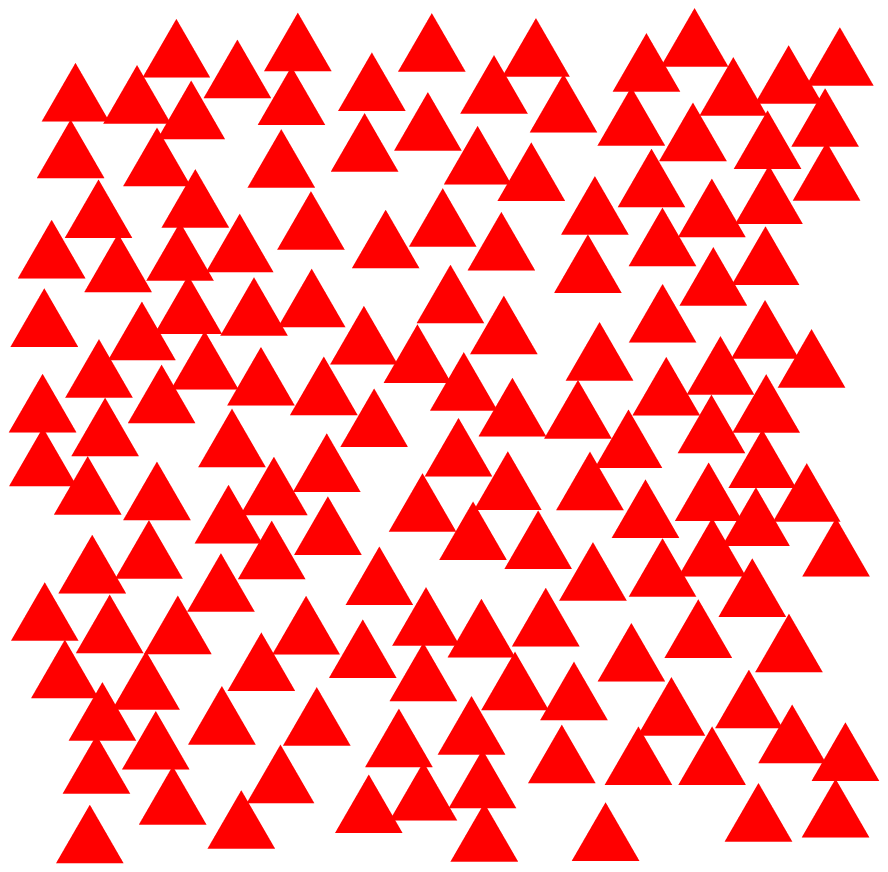}
    \includegraphics[width=0.4\columnwidth]{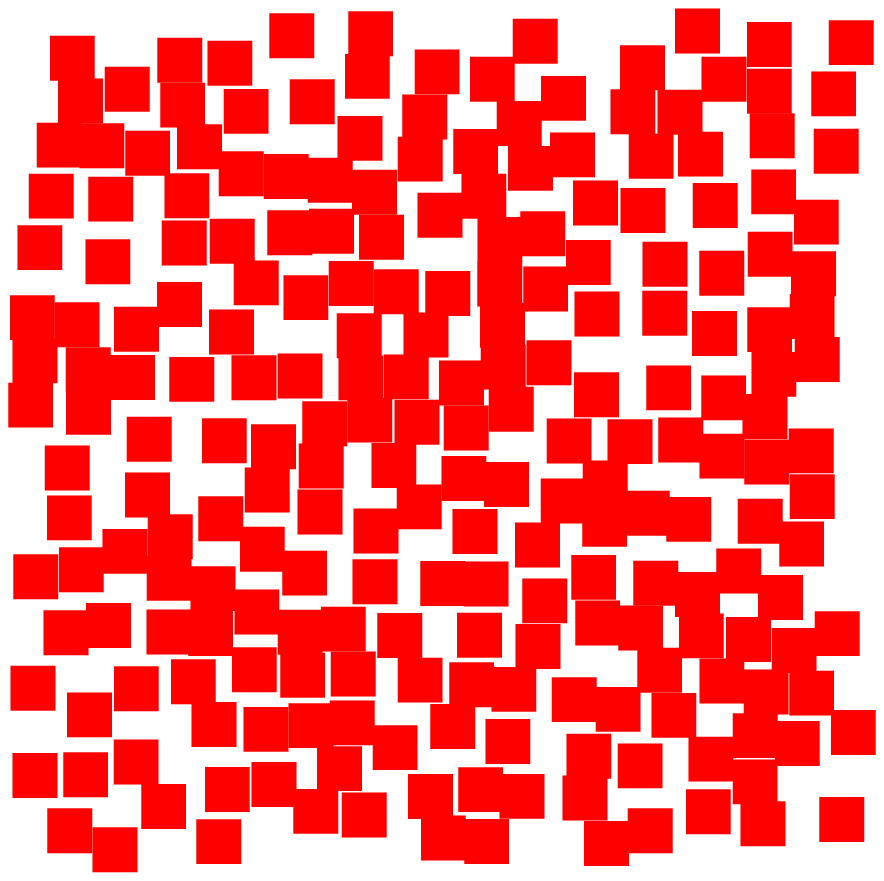}
    \includegraphics[width=0.4\columnwidth]{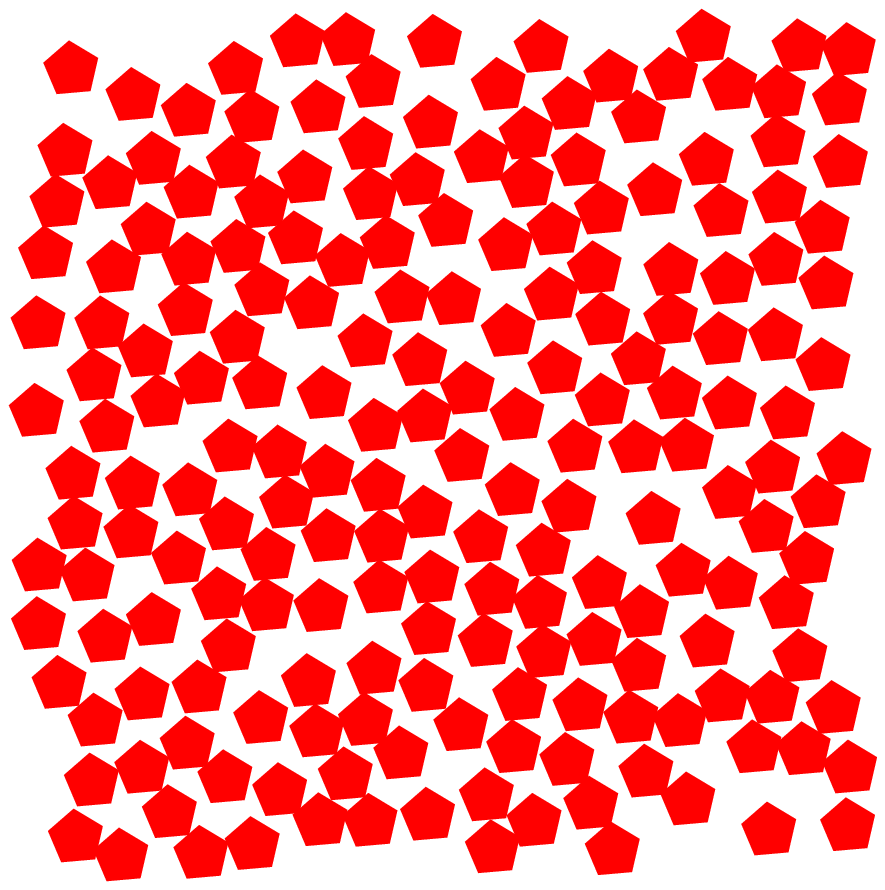}
    \includegraphics[width=0.4\columnwidth]{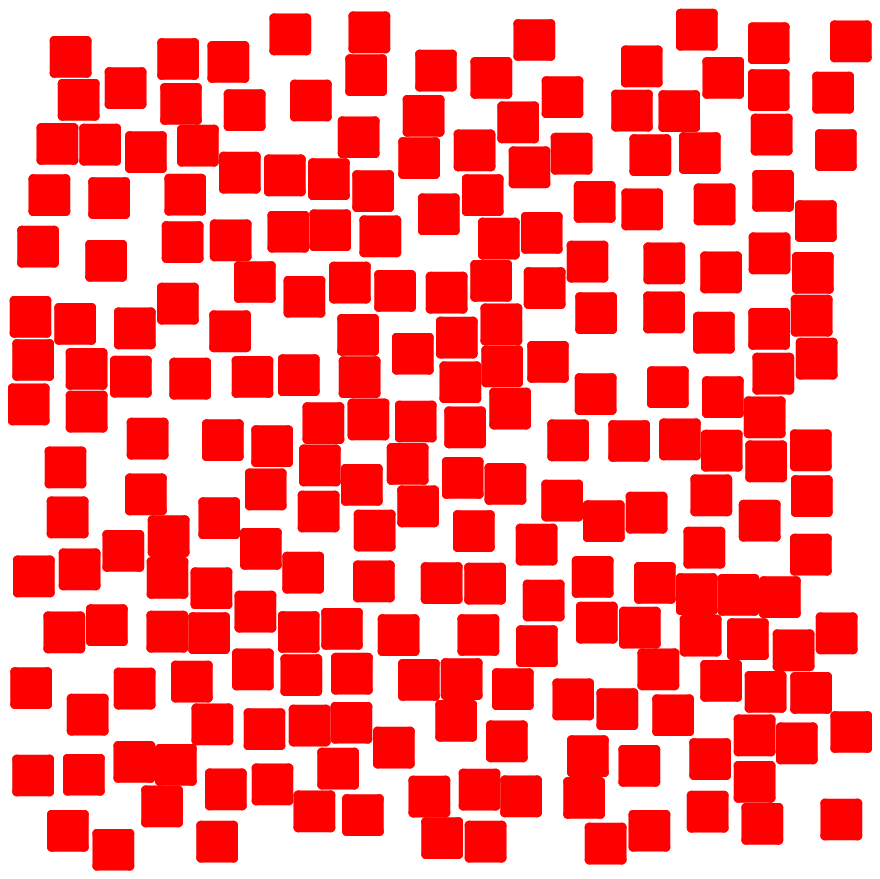}
    \caption{Example saturated packings built of equilateral triangles, squares, pentagons, 
    and rounded squares for $r=0.2$ aligned in parallel. The packing size is $S=400$ and the periodic boundary conditions are used.}
    \label{fig:examples}
\end{figure}

Example saturated packings built of polygons aligned in parallel are shown in Fig.~\ref{fig:examples}.
Note that the square with rounded corners characterized by $r=0.2$ is visually indistinguishable from a \red{normal} square.
\subsection{Kinetics of packing growth}
\begin{figure}[ht]
    \centering
    \includegraphics[width=0.7\columnwidth]{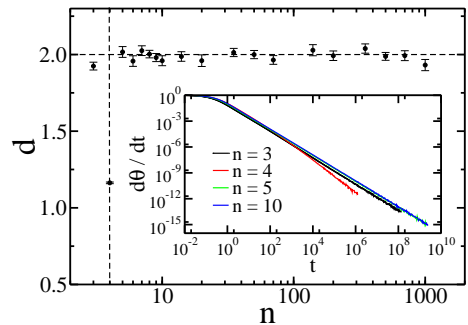}
    \caption{The dependence of the fitted value of the parameter $d$ from (\ref{eq:fl}) on the number of sides $n$ of the polygon. The dashed horizontal line corresponds to $d=2$ and the dashed vertical line denotes $n=4$. Inset shows the kinetics of packings growth for several different regular polygons.}
    \label{fig:d_n}
\end{figure}

\red{The kinetics is presented in Figs.~\ref{fig:kinetics_log} and \ref{fig:d_n}. Fig.~\ref{fig:kinetics_log} presents the data according to (\ref{eq:log}). 
\begin{figure}[ht]
    \centering
    \includegraphics[width=0.8\columnwidth]{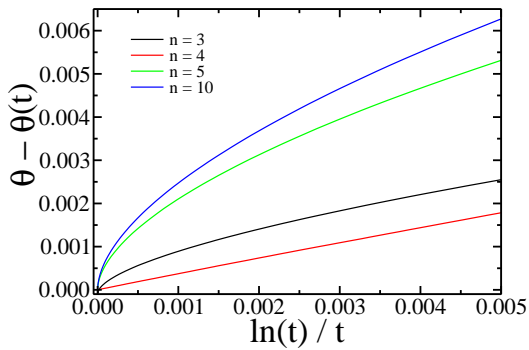}
    \caption{The dependence of the packing fraction near saturation on $\ln (t) / t$ for packing built of oriented regular polygons of a different number of sides. Straight lines correspond \red {to} the kinetics governed by (\ref{eq:log}).}
    \label{fig:kinetics_log}
\end{figure}
The only case where a straight line is observed corresponds to packings built of squares ($n=4$). In analogy, we can analyze the data according to (\ref{eq:fl}).} The dependence of the fitted value of the parameter $d$ from (\ref{eq:fl}) on the number of regular polygon sides $n$ is presented in Fig.~\ref{fig:d_n}. 

We observe that only in the case of packings built of squares the parameter $d$ describing the kinetics of packing growth significantly differs from 2. For all other regular polygons the kinetics seems to be governed by the power law with $d=2$ -- the same as for spheres, which is consistent with the argument that $d$ corresponds to the number of \red{particles’} degrees of freedom \cite{Hinrichsen1986, Ciesla2013}. Here, for all shapes, even for squares, there are only two degrees of freedom corresponding to the position of the center of a two-dimensional shape. For squares, as derived by Swendsen, the kinetics does not follow the power-law (\ref{eq:fl}) but the one described by (\ref{eq:log}) \cite{Swendsen1981}. However, for time scales $t < t_\text{min}$ appearing in our study, it is hard to distinguish between $\log t / t$ and $t^{-\alpha}$ with $\alpha$ slightly smaller than $1$ \red{($d$ slightly larger than $1$)}. Thus, \red{although the RSA kinetics for parallel squares is governed by (\ref{eq:log}) -- see Fig.~\ref{fig:kinetics_log}, we can successfully fit the power law to it -- see Fig.~\ref{fig:d_n}}, with the fitted value of parameter $d$ slightly larger than $1$.

\begin{figure}[t]
    \centering
    \includegraphics[width=0.7\columnwidth]{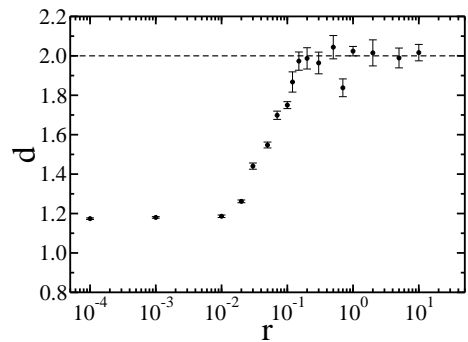}
    \caption{The dependence of \red{the} fitted value of the parameter $d$ from (\ref{eq:fl}) on the parameter $r$ describing rounded squares. The dashed line corresponds to $d=2$.}
    \label{fig:square}
\end{figure}

The above results show the uniqueness of the square shape. This is the only regular polygon that leads to dissimilar kinetics of RSA packing growth. As noted before, to study this phenomenon carefully we also analyzed packings built of squares with rounded corners. The results are presented in Fig.~\ref{fig:square}.
Here, we observe the transition of $d$ for $r$ between $r=0.02$ and $r=0.12$ from $d=1.2$ to $d=2.0$, respectively. Note, that it is very hard to visually distinguish a square from a rounded square even with not insignificant rounding $r=0.2$ -- see Fig.~\ref{fig:examples}. It implies that even tiny changes in shape can significantly influence the RSA kinetics. Similar effects were studied analytically by Baule for two-dimensional shapes placed on a one-dimensional line \cite{Baule2017} and then supported by numerical simulations \cite{Ciesla2020b}. There, the kinetics depended on the analytical nature of the contact function, which is defined as the separation distance at which two particles are in contact. \red{Interestingly, fine details of the contact function are revealed in numerical simulations only for some percentage of packings, and the value of this percentage depends on the packing size \cite{Ciesla2020b}. Therefore, the sharpness and place of the observed transition depend on the packing size.} Regardless, it does not explain the uniqueness of the square shape in comparison with other regular polygons. 

\begin{figure}
    \centering
    \includegraphics[width=0.7\columnwidth]{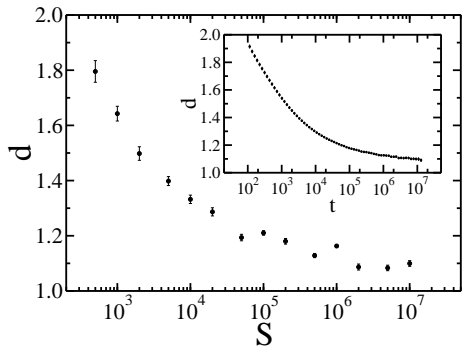}
    \caption{The dependence of the parameter $d$ from (\ref{eq:fl}) on the packing size $S$. Inset shows the dependence of the parameter $d$ on the dimensionless time $t$ at which the packing generation was stopped for packings of size $S=10^7$.}
    \label{fig:d_size}
\end{figure}

The last thing related to the kinetics of packing growth that we want to explain is the recent result obtained in \cite{Moud2022examination}, where the kinetics of RSA of squares is described by the power-law (\ref{eq:fl}) with $d\approx 2$. The packing sizes under consideration in the study mentioned are significantly smaller than the ones we used. Additionally, the author worked with non-saturated configurations. As those two differences may be the source of the discrepancy, we analyzed how the exponent in a power law fit depends on the packing size and the dimensionless time at which we calculate it. The results are presented in Fig.~\ref{fig:d_size}.
The plots clearly \red {show} that the fitted value of the parameter $d$ is larger for both, small and non-saturated packings, which explains the results from the former manuscript. Interestingly, \red{although} the saturated packing fractions can be determined quite accurately using relatively small packings \cite{Ciesla2018}, the study of the kinetics of packing growth requires a few orders of magnitude larger packing sizes. This observation agrees with the results for the kinetics of packing growth for several different figures placed on a one-dimensional line \cite{Ciesla2020b}. It is however important to recall that the value of $d$ for squares will always depend on $t$, regardless of how large it is, because the true asymptotic \red{behavior} is not described by the power law (\ref{eq:fl}), but (\ref{eq:log}).

\begin{figure}
    \centering
    \includegraphics[width=0.7\columnwidth]{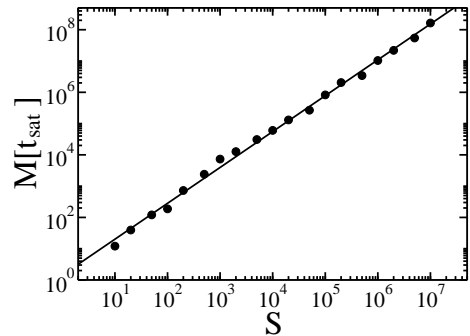}
    \caption{The dependence of the median of saturation time $t_\text{sat}$ on the packing size $S$. Dots are the data determined numerically using $100$ independently generated packing and solid line is a fit $M[t_\text{sat}] = 1.4351 \cdot S^{1.147}$.}
    \label{fig:median}
\end{figure}

Having obtained saturated packing of squares of different sizes we are able to use another way to determine the kinetics of packing growth. It was analytically shown that for disks the median of dimensionless time at which the last shape is added to the packing $M[t_\text{sat}]$ scales with a packing size $S$ as 
\begin{equation}
    M[t_\text{sat}] \sim  S^d,
\end{equation}
where $d$ is the same parameter as in (\ref{eq:fl}) \cite{Ciesla2017}.
It seems that similar relation is also valid for a packing built of oriented squares -- see Fig.~\ref{fig:median}.
Moreover, the \red{fitted} value of the exponent \red{$1.147 \pm 0.016$} is close to the parameter $d$ determined from (\ref{eq:fl}) for $S=10^7$, but here the value is size independent.

In the next sections, we study other basic characteristics of random packings to see if they also reflect the variability of the kinetics of packing growth. 
\subsection{Mean saturated packing fraction}
\begin{figure}
    \centering
    \includegraphics[width=0.7\columnwidth]{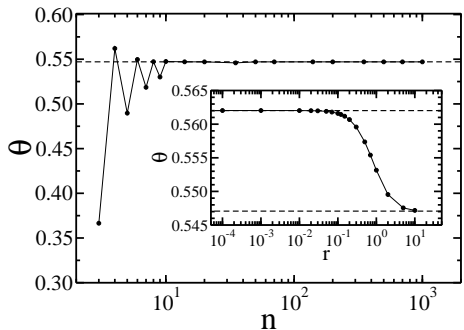}
    \caption{The dependence of \red{the} packing fraction on the number of regular polygon sides. The dashed line corresponds to \red{$\theta = 0.547$ approximating} the RSA packing fraction of disks \cite{Kubala2022, Feder1980, Akeda1975}. The inset shows the dependence of the packing fraction of the rounded square on parameter $r$. The dashed lines highlight two limits: \red{$\theta = 0.547$} for the RSA of disks and \red{$\theta = 0.562$} for the RSA of squares. In both plots, dots are the values determined from generated packings. The error bars are smaller than the dot \red{size and thus they are omitted}. The thin solid lines connecting dots are to guide the eye.}
    \label{fig:q_n}
\end{figure}

The mean density of saturated packing is a basic property \red{of interest}. Here, because we generated strictly saturated configurations, we only need to average \red{the} obtained \red{densities} without any extrapolation. Because the surface area of a single \red{figure} is always normalized to $1$ the density equals the number of deposited shapes divided by \red{the packing area} $S=10^6$. \red{The results obtained} are shown in Fig.~\ref{fig:q_n}.
For regular polygons, we see oscillations of packing fractions for even and odd numbers of polygon sides. This effect was already observed in previous studies \cite{Zhang2018, Moud2022examination}. The values \red{presented} here are, in general, in agreement with these results -- note that in Ref.~\cite{Zhang2018} RSA of unoriented polygons was studied. The packing fraction for rounded squares shows the transition between two limits -- the upper one for aligned squares and the lower one for disks. However, this transition occurs \red{for larger r -- it starts at $r=0.1$ and approaches the packing fraction of disks near $r=1$, while the kinetics for rounded squares is indistinguishable from the one for disks at $r\approx 0.1$}. It \red{shows} that the behavior of packing fractions \red{weakly correlated with} the kinetics of packings \red{for the systems in question}. For convenience, the presented data has been collected in Tab.~\ref{tab:data}.
\begin{table}[ht]
    \centering
    \begin{tabular}{ccc}
        \hline
            $n$ & $\theta$ & $d$ \\
        \hline
            ~~~~$3$~~~~ &	~~~~$0.366410 \pm 0.000016$~~~~	& ~~~~$1.926 \pm 0.025$~~~~ \\
            $4$ &	$0.5620219 \pm 0.0000072$	& $1.100 \pm 0.010$ \\
            $5$	&   $0.489682 \pm 0.000016$	& $2.017 \pm 0.035$	\\
            $6$	&   $0.549713 \pm 0.000016$	& $1.957 \pm 0.035$ \\
            $7$ &	$0.518584 \pm 0.000019$	& $2.025 \pm 0.032$	\\
            $8$ &	$0.547189 \pm 0.000017$	& $2.003 \pm 0.025$ \\
            $9$	&   $0.530072 \pm 0.000018$	& $1.980 \pm 0.026$ \\
            $10$ &	$0.547463 \pm 0.000019$	& $1.960 \pm 0.033$ \\
            $14$ &	$0.547073 \pm 0.000018$	& $1.987 \pm 0.031$ \\
            $20$ &	$0.547049 \pm 0.000018$	& $1.960 \pm 0.038$ \\
            $35$ & 	$0.545938 \pm 0.000018$	& $2.013 \pm 0.027$ \\
            $50$ &	$0.547039 \pm 0.000018$	& $2.000 \pm 0.027$ \\
            $70$ &	$0.547037 \pm 0.000018$	& $1.965 \pm 0.028$ \\
            $140$ &	$0.547037 \pm 0.000018$	& $2.028 \pm 0.036$ \\
            $200$ &	$0.547037 \pm 0.000018$	& $1.992 \pm 0.031$ \\
            $350$ &	$0.547035 \pm 0.000018$	& $2.039 \pm 0.030$ \\
            $500$ &	$0.547035 \pm 0.000018$	& $1.988 \pm 0.026$ \\
            $700$ &	$0.547036 \pm 0.000018$	& $1.993 \pm 0.032$ \\
            $1000$ &$0.547035 \pm 0.000018$	& $1.931 \pm 0.036$ \\
        \hline
            $r$ & $\theta$ & $d$ \\
        \hline
            $0.0001$ & $0.562032 \pm 0.000021$	& $1.1740 \pm 0.0041$ \\
            $0.001$ &  $0.562031 \pm 0.000022$	& $1.1803 \pm 0.0040$ \\
            $0.01$ &   $0.562026 \pm 0.000021$	& $1.1861 \pm 0.0063$ \\
            $0.02$ &   $0.562007 \pm 0.000021$	& $1.2618 \pm 0.0070$ \\
            $0.03$ &   $0.561985 \pm 0.000022$	& $1.441 \pm 0.016$ \\
            $0.05$ &   $0.561909 \pm 0.000021$	& $1.548 \pm 0.015$ \\
            $0.07$ &   $0.561803 \pm 0.000022$	& $1.699 \pm 0.021$ \\
            $0.1$ &    $0.561600 \pm 0.000021$	& $1.750 \pm 0.018$ \\
            $0.12$ &   $0.561448 \pm 0.000022$	& $1.867 \pm 0.051$ \\
            $0.15$ &   $0.561184 \pm 0.000022$	& $1.973 \pm 0.046$ \\
            $0.2$ &    $0.560686 \pm 0.000024$	& $1.987 \pm 0.054$ \\
            $0.3$ &    $0.559581 \pm 0.000025$	& $1.964 \pm 0.055$ \\
            $0.5$ &    $0.557365 \pm 0.000023$	& $2.044 \pm 0.059$ \\
            $0.7$ &    $0.555407 \pm 0.000021$	& $1.838 \pm 0.045$ \\
            $1.0$ &    $0.553160 \pm 0.000019$	& $2.024 \pm 0.024$ \\
            $2.0$ &    $0.549539 \pm 0.000018$	& $2.015 \pm 0.066$ \\
            $5.0$ &    $0.547549 \pm 0.000016$	& $1.989 \pm 0.050$ \\
            $10.0$ &   $0.547177 \pm 0.000017$	& $2.016 \pm 0.042$ \\
        \hline
    \end{tabular}
    \caption{Mean saturated packing fractions obtained from computer simulations. The error of packing fraction $\theta$ is the standard deviation of the mean value. The error of the parameter $d$ was calculated using the exact differential method applied to the result of the least square fitting of numerical data to relation (\ref{eq:fl}).}
    \label{tab:data}
\end{table}
\subsection{Density autocorrelation function}
\begin{figure}
    \centering
    \includegraphics[width=0.8\columnwidth]{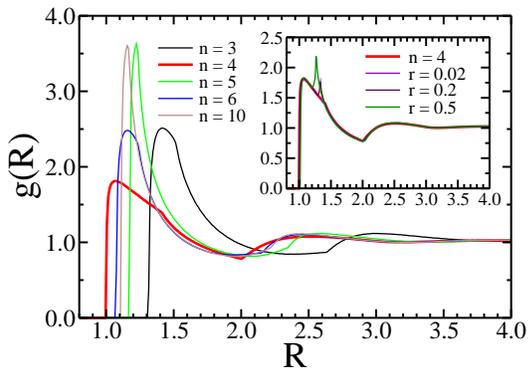}
    \caption{The density autocorrelation function for several packings. The main panel shows the density autocorrelation function for packings built of regular polygons of $n=3, 4, 5, 6$, and $10$ sides. Inset shows the same function but for packings built of rounded squares of $r=0.02$, $0.2$, and $0.5$.}
    \label{fig:correlations}
\end{figure}

While the packing fraction describes the global structure of a set of shapes, the local statistics of their positions can be better understood by probing the density autocorrelation function which can be defined as follows:
\begin{equation}
    g(R) = \lim_{\text{d}R\to 0} \frac{\langle N(R, R+\text{d}R) \rangle}{\theta \, 2 \pi R\, \text{d}R }
\end{equation}
where $\langle N(R, R+\text{d}R) \rangle $ is the mean number of shapes, whose centers are placed in the distance between $R$ and $R + \text{d}R$ from the center of a given figure. The presence of $\theta$ in the denominator is for normalization $g(R\to \infty) = 1$. The density autocorrelation functions for several packings are shown in Fig.~\ref{fig:correlations}.
The correlation functions have typical features of ones observed for RSA packings or equilibrium liquids. It was shown that for one-dimensional packings $g(R)$ vanishes superexponentially \cite{Bonnier1994}, which is also observed here. The plots partially explain the behavior of the mean saturated packing fraction. It is the highest for squares, while at the same time, we observe the shortest distance between neighboring shapes of this type. On the other hand, the plot farthest to the right corresponds to triangles, which form looser configurations. The density autocorrelation for packings built of rounded squares is practically the same as for squares if the rounding is small $r<0.1$. For larger $r$ we observe a second maximum, which grows as the shape approaches the disk. \red{This maximum first appears at $R=\sqrt{2}$, which corresponds to the slight \red {cusp} in $G(R)$ for squares and appears due to its rapid decay when squares are not in touch. However, where square corners are rounded, this distance decreases, and the \red {cusp} transforms into the peak, which travels left with an increasing radius of rounding $r$ and grows up to infinity in the limit of touching disks \cite{Swendsen1981,Pomeau1980}.} The effect of rounding the squares on the positioning of the density autocorrelation function peak could be interesting to study because it would reveal additional factors that could be used to customize the growth kinetics, saturation, and tightness of packing. It is intriguing that while both techniques of going from square to circle finally produced similar d and saturation densities, their response behavior in terms of the density autocorrelation function appears to be different.

The results are \red{expected} because, if these polygons were stacked in a lattice pattern, one would expect squares to be the most densely packed (due to a higher likelihood that the packing would have no corners that could potentially leave some of the available space open). Moreover, particles with odd numbers of sides will undoubtedly have more unoccupied space as their \red{nearest neighbors} cannot occupy \red{the space near them without overlap}. When the number of sides \red{increases}, the maximum saturation packing for the \red{disks} should be reached. A regular polygon's shape may also affect the volume that \red{it excludes}. \red{Because it is more rounded and has a bigger internal volume than the pentagon in respect to its circumference, the hexagon has a lower excluded volume than a regular pentagon of the same size.}
\section{Conclusions}
The square appears to be a unique shape in terms of random sequential adsorption as the two-dimensional oriented packings built of particles of this shape characterize significantly different kinetics given by relation (\ref{eq:log}) while packings built of all other regular polygons, as well as the majority of other shapes obey the power-law (\ref{eq:fl}). By studying the kinetics of \red{packings} built of rounded squares we show that even quite small rounding, which, in practice, is not noticeable visually, changes the kinetics to the one typically observed in similar settings, namely $d=2$ for shapes with two degrees of freedom. This transition is not observed in other characteristics. It is important to add that to study the asymptotic properties of the kinetics of packing growth relatively large packings have to be generated, and preferably as close as possible to their saturation points, contrary to the packing fraction which can be quite precisely estimated using relatively small packings, as long as periodic boundary conditions are used \cite{Ciesla2018}. 

For rounded squares, we also observe the transition between the packing fractions of configurations formed by squares and disks. However, the transition occurs for significantly larger values of parameter $r$ responsible for the amount of rounding than in the case of packing growth kinetics. The study of density autocorrelation functions seems to give additional details regarding packing densities with, \red{however}, no further insight into the asymptotic behavior.

To handle processing in real applications, such as ``\red{Pickering} emulsion'' and adsorption in \red{catalysts}, unique ``particle engineering schemes'' are becoming necessary. For instance, it has been discovered here that for adsorption purposes, items that at a later time can be imagined as molecules or particles with square shapes have distinct growth kinetics and saturation, allowing for customizable levels of adsorbability. Moreover, rounding techniques and modifications made to the same methodology may be adjusted to solve the dearth of theoretical models for \red{two-dimensional} adsorption using RSA. 
\section*{Acknowledgments}
Numerical simulations were carried out with the support of the Interdisciplinary Center for Mathematical and Computational Modeling (ICM) at the University of Warsaw under grant no. GB76-1.
\bibliography{main}
\end{document}